\documentclass{article}
\usepackage{graphicx}
\usepackage{epsfig}

\input{tcilatex}

\begin{document}

\title{\textbf{Critical Behavior of Low Dimensional Magnetic Systems }}
\author{Aycan \"{O}zkan and B\"{u}lent Kutlu* \\
Gazi \"{U}niversitesi, Fen Fak\"{u}ltesi, Fizik B\"{o}l\"{u}m\"{u}, \\
06500 Teknikokullar, Ankara, Turkey.\\
e-mail: aycan@gazi.edu.tr \\
\ *e-mail: bkutlu@gazi.edu.tr}
\maketitle

\begin{abstract}
In this study, critical behavior of low dimensional magnetic systems as
cyano-bridged Tb(III)-Cr(III) bimetallic assembly was investigated with the
mixed spin $3$- spin $3/2$ Ising model. The mixed spin Ising model is
simulated with Cellular Automaton cooling and heating algorithms on
one-dimensional lattices in periodic boundary conditions. The Ising model
Hamiltonian includes only antiferromagnetic nearest-neighbor interaction ($%
J>0$). The mixed spin system behaves like the isolated one-dimensional chain
for zero magnetic field ($h=\frac{H}{J}=0$). In the presence of the magnetic
field, the magnetization is calculated using zero-field cooling ($ZFC$) and
field cooling ($FC$) processes. The one-dimensional Ising model results are
compatible with the cyano-bridged Tb(III)-Cr(III) bimetallic quasi-one
dimensional assembly ( ($\left[ \text{Tb(H}_{2}\text{O)}_{2}\text{(DMF)}%
_{4}\left\{ \text{Cr(CN)}_{6}\right\} \right] \cdot $H$_{2}$O(DMF=
dimethylformamide)) results.

Key words:\textbf{\ }Hysteresis, long-range order,\textbf{\ }Ising model, 
\textbf{\ }cellular automaton.

PACS Numbers: 05.20.-y, 75.10.Hk, 05.10.-a, 75.60.-d.
\end{abstract}

\section{Introduction}

Low dimensional magnetism has been a subject of studies for many years. In
the last decades, new materials have been synthesized to obtain high
temperature magnetism. One of these materials are cyano-bridged $4f$-$3d$
assemblies. $f$-block lanthanide ions having large anisotropic magnetic
moments yield hard magnets and long-range magnetic order in solids $\left[ 4%
\text{, }9\text{-}11\right] $. Some of cyano-bridged $4f$-$3d$ assemblies
also exhibit field-induced magnetic relaxation $\left[ 10\right] $,
cooling-rate dependent magnetism $\left[ 12\right] $, photo-induced
magnetization $\left[ 13\right] $, and humidity response $\left[ 14\right] $
. Interactions between the ions/molecules determine the electronic and
magnetic properties as well as dimensionality of the assembly. Guo et al.
synthesized cyano-bridged Tb (III) -Cr (III) bimetallic assembly ($\left[ 
\text{Tb(H}_{2}\text{O)}_{2}\text{(DMF)}_{4}\left\{ \text{Cr(CN)}%
_{6}\right\} \right] \cdot $H$_{2}$O(DMF= dimethylformamide)) $\left[ 4%
\right] $. They introduced that antiferromagnetic interaction between Tb
(III) and Cr (III) ions represented by $S=$ $3$ and $\sigma =$ $3/2$,
respectively, leads to ferrimagnetic structure in the quasi-one dimensional
zig-zag chain. A transition to 3D longe-range magnetic order from the
ferrimagnetic Tb (III) -Cr (III) chains occurs at $T_{C}=5K$ with the weak
interchain interactions. Therefore, they draw attention to the requirement
of the further experimental and theoretical studies to illuminate the
magnetic interaction mechanism.

The aim of this study was to detect the interaction mechanism of the
one-dimensional spin $3$- spin $3/2$ chain. For this purpose, the
one-dimensional spin $3$ - spin $3/2$ Ising model in its simplest form is
simulated using Cellular Automaton (CA) and the results are compared with
the experimental results to clear up the magnetic interaction mechanism. The
one-dimensional Ising model was first introduced by Ernst Ising in 1925. The
model established by Ising as a chain of spins, each spin interacts only
with its nearest-neighbors, and an external field. At non-zero temperature,
the model does not have any phase transition. Correlation lenght becomes
infinite at $H=T=0$, which is the critical point of the model $[1]$.
However, magnetic order can emerge with broken one-dimensionality due to
orbital degeneracy or quasi-one dimensional geometry $\left[ 2-8\right] $.

The mixed spin Ising model is a simple model to study ferrimagnetism.
Therefore, a variety of spin mixtures, such as spin $1$- spin $1/2$ $\left[
15-22\right] $, spin $1$ - spin $3/2$ $\left[ 23-25\right] $, spin $1$ -
spin $5/2$ $\left[ 26\right] $, spin $2$ - spin $1/2$ $\left[ 15\right] $,
spin $2$ - spin $3/2$ $\left[ 27\text{,}28\right] $, spin $2$ - spin $5/2$ $ %
\left[ 29-34\right] $, spin $1/2$ - spin $3/2$ $\left[ 15\text{,}21\text{,}
35 \right] $, spin $1/2$ - spin $5/2$ $\left[ 15\right] $, spin $3/2$ - spin 
$5/2$ $\left[ 17\right] $, and spin $3$ - spin $3/2$ $\left[ 36\right] $
have been studied frequently by simulation and numerical methods. Creutz
Cellular Automaton (CCA) algorithm and its improved versions are efficient
to study the critical behaviors of the Ising model $\left[ 36-40\right] .$
The CCA algorithm was first introduced by Creutz $\left[ 41\right] $. It is
a microcanonical algorithm interpolating between the conventional Monte
Carlo and the molecular dynamics techniques.

In this study, magnetization ($M$), susceptibility ($\chi $), internal
energy ($U$), and specific heat ($C/k$) are calculated on one-dimensional
chain of linear dimension $L=100$, $500$, $1000$, $5000$, $10000$, $50000$,
and $100000$ with periodic boundary conditions. First, $1D$ behavior and the
long-range order ($LRO$) of the mixed spin system have been investigated
with temperature variation of the thermodynamic quantities in zero external
field ($h=\frac{H}{J}=0$) and external field ($h=\frac{H}{J}\neq 0$) using
the Cellular Automaton cooling algorithm. At the same time, the
thermodynamic quantities are calculated via field cooling ($FC$) and zero
field cooling ($ZFC$) processes for $0\leq h\leq 3.4$. For mixed spin
systems, hysteresis curves are obtained at several temperature values. The
outline of this paper is as follows: In Section 2, the model and the
formalism are given. In Section 3, the results and the discussions are
presented. A conclusion is given in Section 4.

\section{Model}

The mixed-spin Ising model hamiltonian is given by

\begin{equation}
H_{I}=J\sum_{<ij>}S_{i}\sigma _{j}-H\sum_{i}(S_{i}+\sigma _{i})
\end{equation}%
where $S_{i}=0$, $\pm 1$, $\pm 2$ and $\pm 3$ and $\sigma _{j}=$ $\pm 1/2$, $%
\pm 3/2$. $<ij>$ denotes the summation over all nearest -neighbour spin
pairs in a one-dimensional lattice. $J$ is the bilinear interaction ($J>0$)
between $S$ and $\sigma $. $H$ is the external field. The lattice is
established from the two interpenetrating linear chains named as sublattice $%
A$ and sublattice $B$. $S$ and $\sigma $ spins are located in sublattice $A$
and sublattice $B$, respectively (Fig. 1). Three variables are associated
with each site of the lattice. The values of these variables are determined
in each site from its value and those of its nearest- neighbors at the
previous time step. The updating rule, which defines a Cellular Automaton,
is as follows: Of the three variables on each site, the first one is the
Ising spin, $A_{i}$ or $B_{j}$. Its values may be $A_{i}=0$, $1$, $2$, $3$, $%
4$, $5$, and $6$ for $S$ and $B_{j}=0$, $1$, $2$, and $3$ for $\sigma $. $S$
and $\sigma $ can be defined as $S_{i}=(A_{i}-3)$ and $\sigma
_{j}=(2B_{j}-3)/2$ using the Ising spin variables in Eq. (1). The second
variable corresponds the momentum variable which is conjugate to the spin
(the demon). The kinetic energy associated with the demon, $H_{K}$, is an
integer and it is equal to the change in the Ising spin energy ($-dH_{I}$)
for any spin flip.

\begin{equation}
dH_{I}=H_{I}^{t}-H_{I}^{t+1}
\end{equation}%
Kinetic energy values lie in the interval ($0$, $m$) where $m$ takes a
different value for each $h=\frac{H}{J}$. For example, the greatest value of
the $dH_{I}$ equals $-24$ for $J=1$ and  $H=1$. In those terms, $m$ equals $%
48$.

The total energy ($TE$) which is conserved is given in the following form:

\begin{equation}
TE=H_{I}+H_{K}
\end{equation}

The third variable provides a checkerboard row style updating and so it
allows the simulation of the Ising model on a cellular automaton. The black
sites of the checkerboard are updated and then their color is changed into
white; white sites are changed into black without being updated. The
updating rules for the spin and the momentum variables are as follows: For a
site to be updated its spin is changed to one of the other $6$ ($3$) states
with $1/6$ ($1/3$) probability for $S$ ($\sigma $ ) and the change in the
Ising spin energy $dH_{I}$ is calculated. \vspace*{0in}If this energy change
is transferable to or from the momentum variable associated with this site,
such that the total energy $TE$ is conserved, then this change is done and
the momentum is appropriately changed. Otherwise, the spin and the momentum
are not changed. For example, $dH_{I}$ equals $24$ in the case of $\sigma
_{i}^{t}=-\frac{3}{2}$, $S_{i}^{t}=-3$, and $\sigma _{i+1}^{t}=-\frac{3}{2}$
. $S_{i}^{t+1}$can take one of the $S_{i}=$ $3$, $2$, $1$, $0$, $-1$, $-2$,
and $-3$ values at $t+1$ time step \ If the $S_{i}^{t+1}$ takes the value of 
$3$, the $24$ unit energy is transferred to the system as the kinetic energy.

The system temperature for a given total energy is obtained from the average
value of the kinetic energy, which is given by

\begin{equation}
\langle E\rangle =\frac{\sum_{n=0}^{m}ne^{-nJ/kT}}{\sum_{n=0}^{m}e^{-nJ/kT}}
\end{equation}%
where $E=H_{K}$. The expectation value in Eq. (3) is average over the
lattice and the number of time steps. 
\begin{figure}[tbph]
\centering\includegraphics[width=10cm,angle=0,height=10cm]{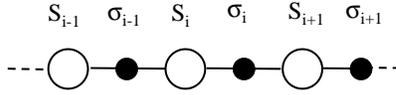}
\caption{One-dimensional lattice in periodic boundary conditions. Sublattice 
$A$ and $B$ generate the one-dimensional lattice. Sublattice $A$ ($B$) is
occupied by $S$ ($\protect\sigma $).}
\end{figure}

The field cooling ($FC$) process and the zero-field cooling ($ZFC$) process
for Tb (III)-Cr (III) are carried out using the cooling and the heating
algorithms of CA $[36-40]$. The cooling and the heating algorithms are
divided into two basic parts, the initialization procedure and the taking of
measurements. In the initialization procedure, firstly, all the spins in the
lattice sites take ferrimagnetic ordered structure ($\uparrow (3)$,$%
^{\downarrow }(-\frac{3}{2})$) and the kinetic energy is given to a certain
percentage of the lattice via the second variables in the black sites such
that the kinetic energy of the site is equal to the change in the Ising spin
energy for any spin flip. The values of the kinetic energy per site is set
to obtain disordered spin configuration for zero field at high temperature.
This configuration is run during the $20000$ cellular automaton time steps.
In the next step, the last configuration in the disordered structure at high
temperature was chosen as a starting configuration for the $FC$ and $ZFC$
simulations. Rather than resetting the starting configuration at each
energy, it was convenient to use the final configuration at a given energy
as the starting point for the next.

\subsection{FC and ZFC Processes}

In the measurement step of the $FC$ algorithm, the last configuration of the
initialization procedure in the disordered structure is taken as a starting
configuration. The spin system is cooled for a value of non-zero field ($h=%
\frac{H}{J}\neq 0$). During the cooling cycle, a certain amount of energy
per site are subtracted from the lattice through the second variable ($H_{K}$%
) after the $2000000$ cellular automaton steps. In the zero-field cooling
process ($ZFC$), the initial configuration in the disordered structure is
used as a starting configuration for the cooling run at zero-field ($h=0$).
The last configuration at low temperature of the cooling process is taken as
a starting configuration for the heating run of the $ZFC$. Then the spin
system is heated for a value of non-zero field ($h=\frac{H}{J}\neq 0$).
During the heating cycle, a certain amount of energy per site is given to
the lattice through the second variable ($H_{K}$) after the $2000000$
cellular automaton steps. These energy amounts are determined considering
the $dH_{I}$ values for the possible spin configurations. Thus, the whole
energy is used by the spin system. As a result, the spin system does not
contain the remnant energy, which affects the temperature measurement.

\section{Results and Discussions}

\begin{figure}[tbph]
\centering\includegraphics[width=12cm,angle=0,height=12cm]{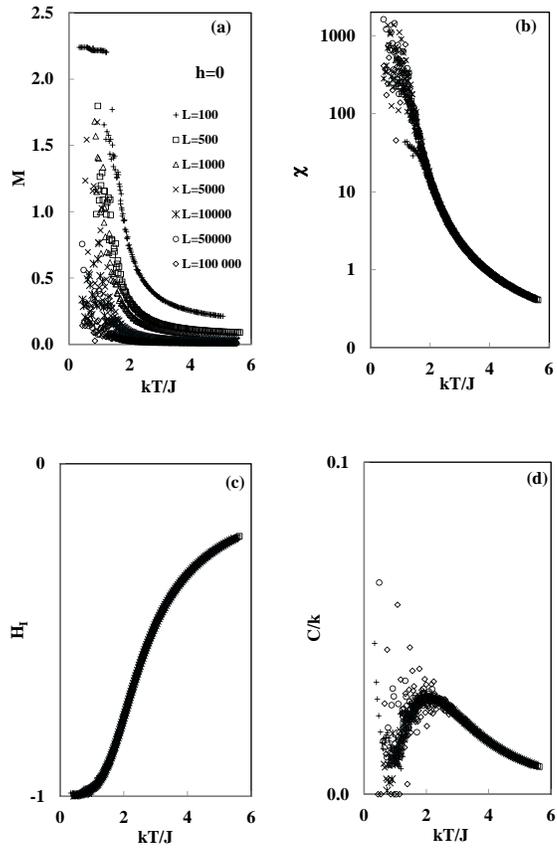}
\caption{Temperature dependence of (a) Magnetic order parameter ($M$), (b)
Susceptibility ($\protect\chi $), (c) Internal energy ($U$), and (d)
Specific heat ($C/k$) at $h=H/J=0$ on $L=100$, $1000$, $5000$, $10000$, $%
50000$, and $L=100000$.}
\end{figure}

All simulations were carried out using the cooling and the heating
algorithms improved from CCA for the one-dimensional spin $3$ - spin $3/2$
Ising model. The thermodynamic quantities (the order parameter ($M$), the
susceptibility ($\chi $), the internal energy ($U$), and the specific heat ($%
C/k$)) were computed over the lattice and over the number of time steps ($%
2000000$ ) after the discard of the first $100000$ time steps during the
development of the cellular automaton. Thus, the values of the thermodynamic
quantities correspond to the equilibrium average values. The calculations
were repeated by field cooling ($FC$) and the zero-field cooling ($ZFC$)
processes on one-dimensional lattices with the linear dimensions $L=100$, $%
500$, $1000$, $5000$, $10000$, $50000$, and $100000$ for periodic boundary
conditions.

The thermodynamic quantities are calculated from

\begin{equation}
M=\frac{1}{N}\sum_{i}S_{i}-\frac{1}{N}\sum_{j}\sigma _{j}
\end{equation}
\begin{equation}
U=\frac{1}{H_{0}}((\sum_{<ij>}S_{i}\sigma _{j})-\frac{H}{J}
\sum_{i}(S_{i}+\sigma _{j}))
\end{equation}

\begin{equation}
\chi =N\frac{\left\langle M^{2}\right\rangle -\left\langle M\right\rangle
^{2}}{kT}
\end{equation}

\begin{equation}
C/k=N\frac{\left\langle U^{2}\right\rangle -\left\langle U\right\rangle ^{2} 
}{(kT)^{2}}
\end{equation}
where $H_{0}$ is the ground state energy at $kT/J=0$.

\subsection{The behavior of the one-dimensional spin system in the absence
of an external magnetic field ($h=\frac{H}{J}=0$)}

The temperature dependence of the order parameters ($M$), the susceptibility
($\chi $), the internal energy ($U$), and the specific heat ($C/k$) in the
absence of magnetic field ($h=0$) were illustrated in Fig. 2 for the cooling
algorithm. As it is seen in Fig. 2(a), the value of $M$ increases with
decreasing temperature ($kT/J\rightarrow 0$) for each lattice size similar
to the one-dimensional Ising model. However, the susceptibility diverges to
infinity at absolute zero temperature for all lattice sizes (Fig. 2(b)). At
the same time, $U$ does not have an inflection indicating any phase
transition and $C/k$ exhibits a broad peak.The critical behavior of $M$, $
\chi $, $U$, and $C/k$ are compatible with the behavior of the
characteristic one-dimensional Ising model, and the isolated one-dimensional
Ising chain results which has very weak interchain coupling in 3-d
dimensional lattice $[42]$.

The evolution of the magnetization for each lattice was plotted in Fig. 3 as
a function of lenght per $200$ sites ($L/200$) for $kT/J=0.8$ at $h=0$. As
it is seen, there are local order regions which are separated by the
fluctuations between pozitive and negative magnetization values for $L\geq
10000$ lattices. The spatial fluctuations decrease with decreasing lattice
size. This causes an increase in the order parameter at low temperatures for
decreasing lattice sizes. Therefore, the one-dimensional Ising model can be
modeled in only large lattice size as $L\geq 10000$. In this study, the
lattice size was selected as $L=100000$ for the simulation of the
one-dimensional Ising model

For $L=100000$, spatial behavior of order parameter are shown in Fig. 4 for $%
kT/J=0.446$, $0.808$, $1.014$, and $1.510$ at $h=0$. At $kT/J>0$, the
magnetization fluctuates between the local regions for all temperature
values. The fluctuations are more often with increasing temperature and so
the local order regions disappear as expected for the one-dimensional Ising
model.

\begin{figure}[tbph]
\centering\includegraphics[width=10cm,angle=0,height=10cm]{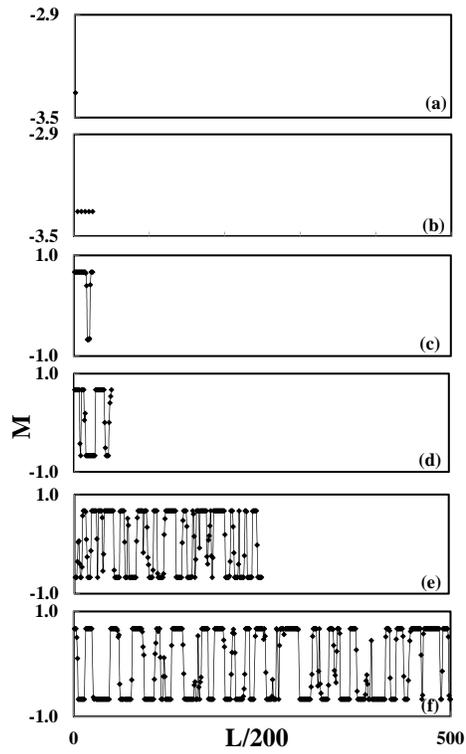}
\caption{Magnetic order parameter as a function of lenght per $200$ sites ($%
L/200$) for $h=\frac{H}{J}=0$ on (a) $L=100$, (b) $L=1000$, (c) $L=5000$,
(d) $L=10000$, (e) $L=50000$, and (f) $L=100000$ at $kT/J=0.8$.}
\end{figure}

\begin{figure}[tbph]
\centering\includegraphics[width=10cm,angle=0,height=10cm]{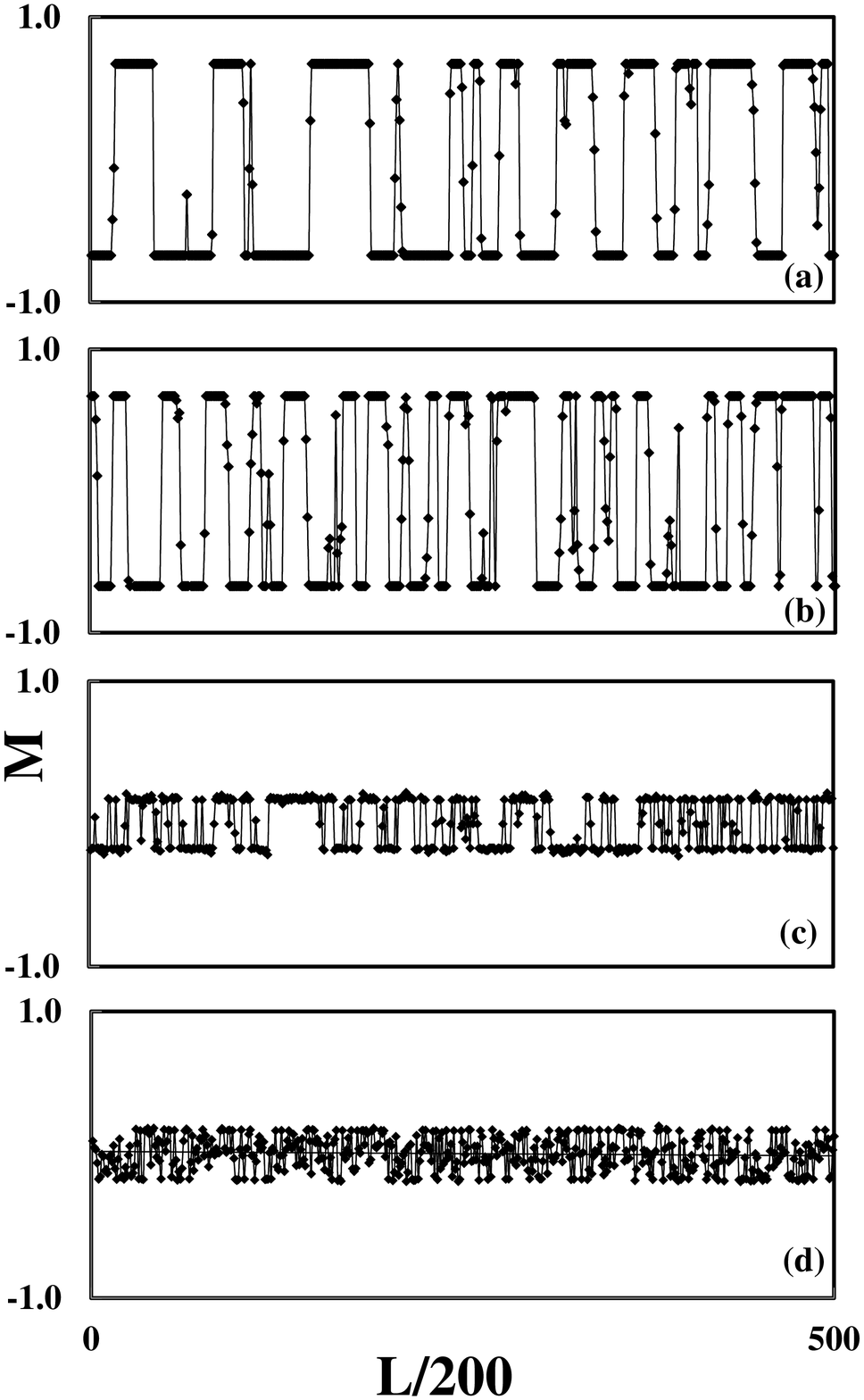}
\caption{Magnetic order parameter as a function of lenght per $200$ sites ($%
L/200$) for $h=\frac{H}{J}=0$ at (a) $kT/J=0.446$, (b) $kT/J=0.808$, (c) $%
kT/J=1.014$ , and (d) $kT/J=1.510$ on $L=100000$.}
\end{figure}

\subsection{The behavior of the one-dimensional spin system in an external
magnetic field ($h=\frac{H}{J}\neq 0$)}

For $h=0.1$, the evolution of the magnetization was plotted as a function of
lenght per $200$ sites ($L/200$) for different temperatures ($kT/J=0.439$, $
2.264$, $2.499$ and $3.513$) in Fig. 5. As it is seen, the local order
regions do not occur through the lattice for $kT/J\neq 0$. For $kT/J<2.499$,
the system shows a long range order ($LRO$) with the effect of the external
magnetic field. On the other hand, the $LRO$ begins to decay at a critical
temperature and disappears at high temperatures. This behavior indicates a
phase transition depending on the temperature. The temperature dependence of
the magnetization was obtained by field cooling ($FC$ ) process and zero
field cooling ($ZFC$) process in the interval $0\leq h\leq 3.4$. \smallskip

In Fig. 6(a) and 6(b), the temperature variation of $M$ and $\chi $ are
illustrated for $ZFC$ and $FC$ processes at $h=0.4$. \ The values of
magnetization for $ZFC$ and $FC$ are different than each other at low
temperatures. This difference disappears at high temperatures (Fig. 6(a)).
The susceptibilities obtained with $ZFC$ and $FC$ processes have a peak at
the same temperature (Fig. 6(b)). The simulation results are in agreement
with the experimental cyano-bridged terbium (III)-chromium (III) bimetallic
quasi-one dimensional assembly result $\left[ 4\right] $. In Fig. 6(c), the
temperature variations of the magnetizations for $ZFC$ and $FC$ are shown
for several magnetic field values. As it seen in the figure, the difference
between magnetizations of the $FC$ and the $ZFC$ disappears with increasing $%
h$ value at low temperatures. \ Using the $ZFC$ and $FC$ magnetization
values, the hysteresis curves were obtained for $L=100000$. In Fig. 7(a) and
7(b), hysteresis curves are illustrated for $kT/J=1.5$ and $2$ temperature
values. As it is seen from hysteresis curves, the spin system has remnant
magnetization when the magnetic field drops to zero. The remnant
magnetization values are estimated from the hysteresis curves for several
temperatures and shown in Fig. 7(c). The temperature variation of remnant
magnetization is similar to spontaneous magnetization of a magnetic system
above one dimension.

\begin{figure}[tbph]
\centering\includegraphics[width=10cm,angle=0,height=10cm]{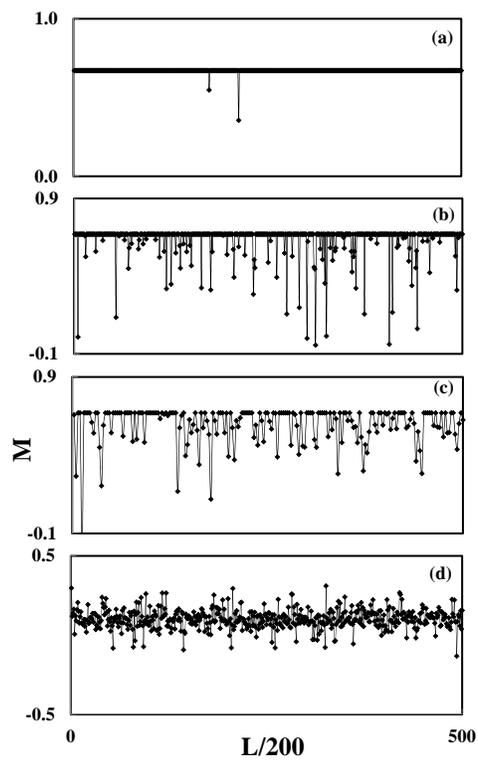}
\caption{Magnetic order parameter as a function of lenght per $200$ sites ($%
L/200$) for $h=\frac{H}{J}=0.1$ at (a) $kT/J=0.439$, (b) $kT/J=2.264$, (c) $%
kT/J=2.499$ and (d) $kT/J=3.513$ on $L=100000$.}
\end{figure}

\begin{figure}[tbph]
\centering\includegraphics[width=10cm,angle=0,height=10cm]{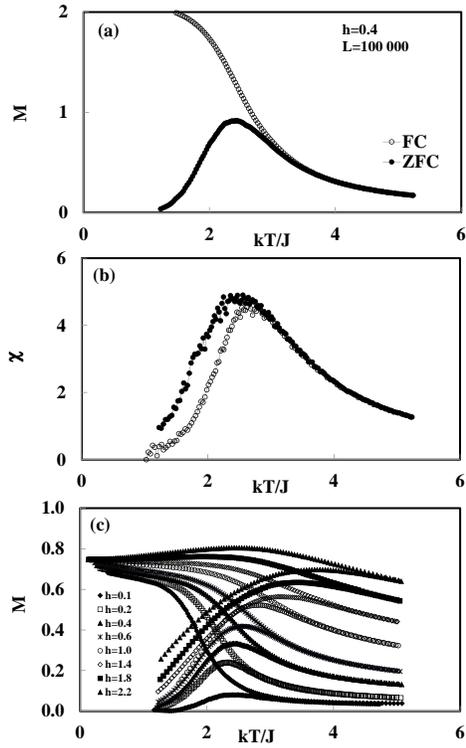}
\caption{The temperature dependences of (a) the magnetization ($M$), and (b)
the susceptibility ($\protect\chi $) for $h=\frac{H}{J}=0.4$. (c) The
temperature dependences of the magnetization ($M$) in the interval $0\leq
h\leq 3.4$. The thermodynamic quantities are obtained by $FC$ process and $%
ZFC$ process on $L=100000$.}
\end{figure}

\begin{figure}[tbph]
\centering\includegraphics[width=10cm,angle=0,height=10cm]{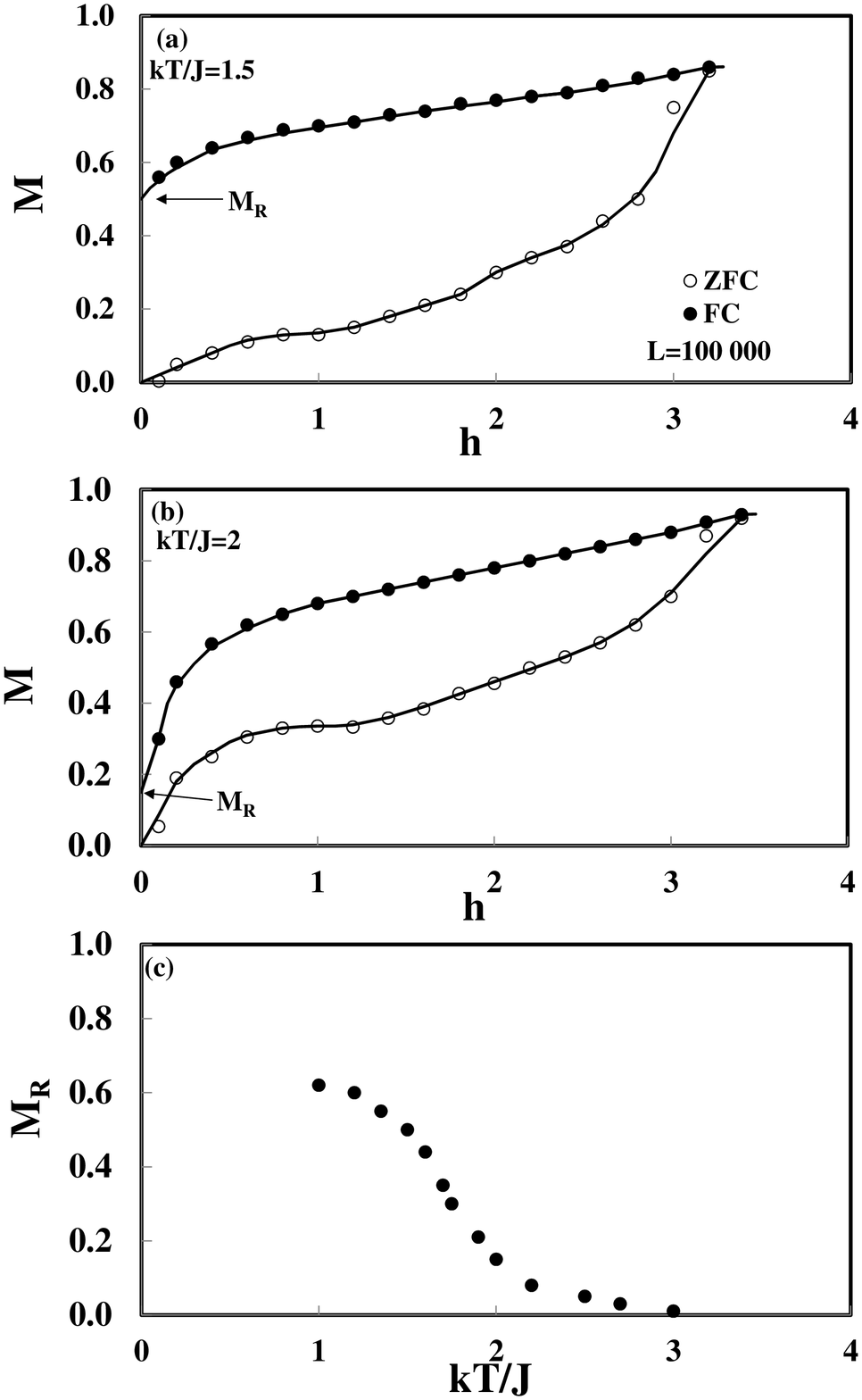}
\caption{The hysteresis curves obtained using the zero field cooling ($ZFC$
) and the field cooling ($FC$) magnetization measurements on $L=100000$ at
(a) $kT/J=1.5$, (b) $kT/J=2.0$, and (c) The temperature dependences of
remnant magnetization ($M_{R}$).}
\end{figure}

\section{Conclusion}

The mixed spin $3$- spin $3/2$ Ising model with antiferromagnetic
nearest-neighbor interaction is simulated on one-dimensional lattices with
linear dimension $L=100$, $500$, $1000$, $5000$, $10000$, $50000$, and $%
100000$ using Cellular Automaton cooling and heating algorithms improved
from Creutz Cellular Automaton (CCA). The values of order parameter ($M$),
susceptibility ($\chi $), Internal energy ($U$) and specific heat ($C/k_{B}$
) are calculated using cooling and heating algorithms ($h=\frac{H}{J}=0$).
The mixed spin system shows the one-dimensional Ising chain behavior for $h=0
$ $\left[ 42\right] $. At the same time, the system exhibits a long-range
order ($LRO$ ) at low temperatures on the one-dimensional lattice in the
presence of the external magnetic field ($h=\frac{H}{J}=0.1$). The
hysteresis curves are obtained from zero field cooling ($ZFC$) and field
cooling ($FC$) magnetization values in the presence of the external magnetic
field ($h\neq 0$). It is seen that the mixed spin $3$- spin $3/2$ Ising
model has remarkable remnant magnetization as presented in the experimental
study $\left[ 4\right] $. The results of the one-dimensional Ising model
with antiferromagnetic nearest-neighbor interaction are similar with
cyano-bridged Tb (III) -Cr (III) bimetallic quasi-one dimensional assembly ($%
\left[ \text{Tb(H}_{2}\text{O)}_{2}\text{(DMF)}_{4}\left\{ \text{Cr(CN)}%
_{6}\right\} \right] \cdot $H$_{2}$O(DMF= dimethylformamide)) results $\left[
4\right] $. As a result, the magnetic behavior of the cyano-bridged Tb (III)
-Cr (III) bimetallic assembly can be determined by intrachain
nearest-neighbor interactions in the absence of interchain interaction on
one-dimension. Thus, the cyano-bridged Tb (III) -Cr (III) bimetallic
assembly may be considered as one dimensional instead of quasi one
dimensional spin system. Our calculations show that the high temperature
phase transition on a one-dimensional mixed spin $3$- spin $3/2$ system,
similar to the cyano-bridged Tb (III) -Cr (III) bimetallic assembly,\ is
caused by the ferrimagnetic nature of the spin system.

\section{References}

$\left[ 1\right] $ R.J. Baxter, \textit{Exactly solved models in statistical
mechanics} (Academic press inc., San Diego, 1989).

$\left[ 2\right] $ K.S. Asha, K.M. Ranjith, A. Yogi, R. Nath, S. Mandal,
Dalton Trans. \textbf{44}, 19812 (2015).

$\left[ 3\right] $ J. M. Law, H.-J. Koo, M.-H. Whangbo, E. Br\"{u}cher, V.
Pomjakushin, and R. K. Kremer, Phys. Rev. B \textbf{89}, 014423 (2014).

$\left[ 4\right] $ Y. Guo, G.-F. Xu, C. Wang, T.-T. Cao, J. Tang, Z.-Q. Liu,
Y. Ma, S.-P. Yan, P. Cheng, D.-Z. Liao, Dalton Trans. \textbf{41}, 1624
(2011).

$\left[ 5\right] $ X. Zhang, G. Zhang, T. Jia, Y. Guo, Z. Zeng, H.Q. Lin,
Phys. Lett. A \textbf{375}, 2456 (2011).

$\left[ 6\right] $ S. Shiraki, J of Vacuum Soc. of Japan \textbf{52}, 595
(2009).

$\left[ 7\right] $ M. Estrader, J. Ribas, V. Tangoulis, X. Solans,
M.Font-Bardia, M. Maestro, C. Diaz, Inorg. Chem. \textbf{45}, 8239 (2006).

$\left[ 8\right] $ A. Figuerola, C. Diaz, M.S.El Fallah, J. Ribas, M.
Maestro, J. Mahia, Chem. Commun. \textbf{13}, 1204 (2001).

$\left[ 9\right] $ F. Hulliguer, M. Landolt, H. Vetsch, J. Solid State Chem. 
\textbf{18}, 283 (1976).F. Hulliger, M. Landolt, H. Vetsch

$\left[ 10\right] $ Y.-Z. Zhang, G.-P. Duan, O. Sato, S. Gao, J. Mater.
Chem. \textbf{16}, 2625 (2006).

$\left[ 11\right] $ H. Zhou, A.-H. Yuan, S.-Y. Qian, Y. Song, G.-W. Diao,
Inorg. Chem. \textbf{49}, 5971 (2010).

$\left[ 12\right] $ T. Hozumi, S.-\.{I}. Ohkoshi, Y. Arimoto, H. Seino, Y.
Mizobe, K. Hashimoto, J.\ Phys. Chem. B \textbf{107}, 11571 (2003).

$\left[ 13\right] $ H. Svendsen, J. Overgaard, M. Chevallier, E. Collet,
Y.-S. Chen, F. Jensen, B.B. Iversen, Chem. Eur. J. \textbf{16}, 7215 (2010).

$\left[ 14\right] $ S.-\.{I}. Ohkoshi, K.-\.{I}. Arai, Y. Sato, K.
Hashimoto, Nat. Mater. \textbf{3}, 857 (2004).

$\left[ 15\right] $ A. Dakhama and N. Benayad, J.of Magn. and Magn.Mater. 
\textbf{213}, 117 (2000).

$[16]$ M. Azhari, N. Benayad, M. Mouhib, Superlattices and Microstructures 
\textbf{79}, 96 (2015).

$[17]$ G.M. Buendia and M.A. Novotny, J. Phys. Condens. Matter. \textbf{9},
5951 (1997).

$[18]$ E. Aydiner, Y. Yuksel, E. Kis-Cam, H. Polat, J.of Magn. and
Magn.Mater. \textbf{321}, 3193 (2009).

$[19]$ W. Selke, J. Oitmaa, J.Phys: Condens. Matter. \textbf{22}, 076004
(2010).

$[20]$ F.W.S. Lima, M.A. Sumour, Physica A \textbf{391}, 948 (2012).

$[21]$ A. Zaim and M. Kerouad, Physica A \textbf{389}, 3435 (2010).

$\left[ 22\right] $ E. Albayrak, Solid State Communications \textbf{159}, 76
(2013).

$[23]$ Y. Nakamura, J.W. Tucker, IEEE Transactions on magnetics \textbf{38},
2406 (2002).

$[24]$ A. Bob\'{a}k, O.F. Abubrig and D. Horv\'{a}th, J.of Magn. and Magn.
Mater. \textbf{246}, 177 (2002).

$\left[ 25\right] $ M. Erta\c{s}, M. Keskin, Physics Letters A \textbf{379},
1576 (2015).

$[26]$ B. Deviren, M. Bat\i\ and M. Keskin, Phys. Scr. \textbf{79}, 065006
(2009).

$[27]$ W.Guo-Zhu, M.Hai-Ling, Commun.Theor. Phys. \textbf{51}, 756 (2009).

$[28]$ B. Deviren, Y. Polat, M. Keskin, Chin. Phys. B \textbf{20}, 060507
(2011).

$[29]$ W. Jiang, W. Wang, F. Zhang and W.J. Ren, J.of Appl. Phys. \textbf{\
105 }, 07E321 (2009).

$[30]$ Y.Kai-Lun, L.Jian-Wen, L.Zu-Li, F.Hua-Hua and Z. Lin, Theor. Phys. 
\textbf{47}, 741 (2007).

$[31]$ M. Erta\c{s}, M. Keskin, B. Deviren, Physica A \textbf{391}, 1038
(2012).

$[32]$ S.G. Carling, P. Day, Polyhedron \textbf{20}, 1525 (2001).

$[33]$ Y. Nakamura, J.Phys. Condens. Matter. \textbf{12}, 4067 (2000).

$[34]$ M. Charilaou, K.K. Sahu, A.U. Gehring and J.F. L\"{o}ffler, Phys.
Rev. B \textbf{84}, 224434 (2011).

$[35]$ B. Deviren, M. Keskin, O. Canko, J.of Magn. and Magn.Mater. \textbf{\
321}, 458 (2009).

$[36]$ A. \"{O}zkan, Phase transitions \textbf{89}, 94 (2016).

$\left[ 37\right] $ A. \"{O}zkan, B. Kutlu, Cent. Europ. J. of Phys. \textbf{%
\ \ 9}, 884 (2011).

$\left[ 38\right] $ A. \"{O}zkan, B. Kutlu, Int. J. of Mod. Phys. C \textbf{%
\ \ 20}, 1617 (2009).

$\left[ 39\right] $ B. Kutlu, A. \"{O}zkan, N. Sefero\u{g}lu, A. Solak and
B. Binal, \textit{Int. J. Mod. Phys.C}\textbf{\ 16,} 933 (2005).

$\left[ 40\right] $ N. Sefero\u{g}lu, B. Kutlu, Physica A \textbf{374}, 165
(2007).

$\left[ 41\right] $ M. Creutz, Ann. Phys. \textbf{167}, 62 (1986).

$[42]$ T. Graim, D.P. Landau, Phys. Rev. B \textbf{24}, 5156 (1981).

\end{document}